# ACTIVE OPTICS CONTROL OF THE VST TELESCOPE WITH THE CAN FIELD-BUS

D. Mancini, P. Schipani, G. Mazzola, L. Marty, M. Brescia, F. Cortecchia, F. Perrotta, E.Rossi
Osservatorio Astronomico di Capodimonte, Napoli, Italy

Abstract

The VST (VLT Survey Telescope) is a 2.6 m class Alt-Az telescope to be installed at Mount Paranal in the Atacama desert, Chile, on the European Southern Observatory (ESO) site. The VST is a wide-field imaging facility planned to supply databases for the ESO Very Large Telescope (VLT) science and carry out stand-alone observations in the UV to I spectral range, [1]. This paper will focus on the distributed control system of active optics based on CAN bus and PIC microcontrollers. Both axial and radial pads of the primary mirror will be equipped by astatic lever supports controlled by microcontroller units. The same CAN bus + microcontroller boards approach will be used for the temperature acquisition modules.

## 1 VST ACTIVE OPTICS GENERAL CONCEPTS

The active optics of the VST telescope is based on a complex control scheme; both primary and secondary mirrors must be supported and accurately positioned.

The aberrations corrected by the VST active optics control system are essentially the coma, defocus (secondary mirror re-alignment) and the spherical, astigmatism, quad-astigmatism and tri-coma (primary mirror sag deformation) aberrations, [3].

Hereinafter the description will focus on the primary mirror active optics system.

The VST primary mirror figure is controlled by a set of active supports based on astatic levers. They are driven in order to elastically compensate wave-front deformations measured by a wave-front sensor in a feedback loop. The mirror is supported by means of axial and lateral pads.

VST has 84 axial pads distributed on four concentric rings, which include respectively from the center out, 12, 18, 24 and 30 pads, (figure 1). The axial pad distribution was analyzed by means of Finite Elements Analysis (FEA).

24 lateral pads provide the lateral support. The lateral component of the weight of the mirror is supported by the 24 lateral levers, when the altitude axis of the telescope is moved. The forces are applied at the center of the rim and the force vector lies in the neutral plane at the outer edge.

Each M1 mirror support system must be driven to a required force and this force must be monitored in a feedback loop scheme. The monitoring of the net force developed by each single actuator is obtained from a load-cell. The speed of the single motor is monitored with a tachometer.

The wavefront analysis system is based on a Shack-Hartmann sensor and is sensitive enough to work on guide stars of magnitude +14 for integration times of about 30 seconds. The size of the VST pupil sampling sub-apertures is on the order of 250 mm. The number of the Shack-Hartmann spots across the telescope pupil is equivalent to 10. The lenslet array shows a f/45 focal number and a lenslet diameter $Dl \approx 0.5$ mm.

The Adapter/Rotator of the VST includes a sensing and guiding arm that provides the telescope guiding and the on-line wavefront analysis.

The guiding and sensing arm is coaxially mounted on a rotating support with the rotator and its optics are mounted on a radial rail. These two devices implement a polar system, which allows the arm to reach the center of telescope field during the off-line calibration phases.

The polynomial base used to fit the wavefront is determined by the elastic mode of the primary mirror itself. In this way the energy required to deform the primary mirror is minimized. The deformation of the primary mirror is obtained by variation of the overall pattern of forces applied by the astatic lever supports.

There are three different kinds of corrections to apply to the M1 sag through the astatic levers:

- compensation of the axial component of M1 weight (Passive force)
- correction of the optical distortion of M1 (Passive distortion)
- compensation of the optical distortion changing with altitude angle (Active distortion).

While the passive distortion is constant, both the passive force and active distortion depend on the altitude angle. The sag of the primary mirror can be corrected to eliminate the following aberrations: spherical, astigmatism, triangular coma, quadratic astigmatism and fifth order coma.

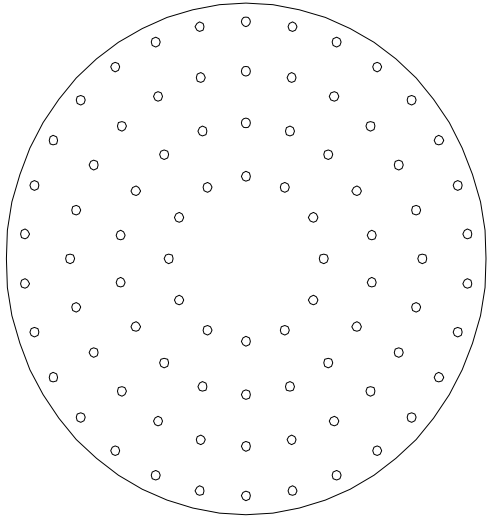

Figure 1: M1 axial pad distribution

## 2  M1 CONTROL CAN BUS

The VST M1 84 axial and 24 radial pads are controlled by a CAN bus network of astatic lever controllers connected to a CAN bus interface board in a Local Control Unit (Figure 2).

The Local Control Unit (LCU) is a VME bus computer equipped with a Motorola MVME 2604 CPU card. The CAN bus interface module is a TEWS Datentechnik TPMC816 2 channel CAN bus PMC module, mounted into the PCI mezzanine socket available on the Motorola CPU card. The CAN bus interface is based on Intel 82527 chipset. This board is the master of the CAN bus network and can address the slave modules (Astatic Lever Controllers) to send configuration parameters or to request data. The address mechanism is based on the acceptance filtering capability. All CAN implementations provide some hardware acceptance filters to relieve the microcontroller from the task of filtering the needed messages from those not of interest one. The address space managed by the acceptance filter is equal to 127 devices, which is sufficient for the VST active optics application, [3].

The LCU runs a VxWorks real time operating system. The standard ESO architecture is adopted both for the hardware and software architecture. The LCU running the active optics pad control software is one of the nodes of the overall VST telescope control network. This is composed by higher level coordination and control HP-Ux workstations and low level LCUs to directly interface with the controlled devices.

The physical layer is based on a twisted pair for data transmission. Two other wires are present to distribute 12V power to all modules connected to the network. The maximum data rate envisaged for the CAN bus is 1 Mbit/s.

The power for the slave modules electronics is provided by the CAN bus; one DC/DC converter for each module performs 12V/5V conversion. The power supply for the actuators, sensors and the electronics connected after the insulation barrier is distributed by a field power bus.

The slave modules contain a microcontroller that provides for communication protocol management and control of the hardware devices present on the module. The operational parameters are communicated from the LCU to the slave module, which locally implements the closed loop control without any other intervention of the LCU.

The driving of the array of actuators using a distributed system of local microcontrollers has some advantages, such as quick positioning loop, in place, with the highest allowed bandwidth, not limited by intercommunication mechanisms, [1].

A custom design was chosen for the astatic lever controller, because no commercial solution able to completely fulfill the application requirements was available.

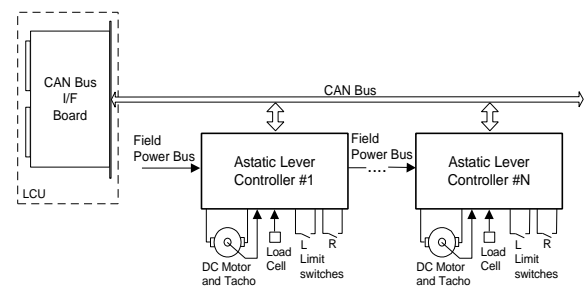

Figure 2: CANbus network configuration

The functional requirements for the Astatic Lever Controller are:

- Control of the applied force by linear regulators
- Control of the DC motor
- Acquisition of force from the load cell
- Acquisition of the end of travel switches status
- Acquisition of motor speed from the tachometer
- Implementation of the physical and data link layer according the ISO 11898

- Implementation of the data link layer according to the CAN 2.0 B standard
- Implementation of monitoring and debugging functions via CAN bus and local RS232 serial port
- Galvanic insulation of motors and sensors by opto-couplers and insulated DC/DC converters

Components of the Astatic Lever Controller are:

- Microcontroller
- CAN bus communication controller
- CAN bus interface drivers
- Opto-insulated Load Cell Interface
- Opto-Insulated Tachometer Interface
- Opto-Insulated DC Motor Driver
- Software Limit Switch Interface
- RS232 serial port for monitoring and debugging functions

The microcontroller platform selected to implement the custom modules is the PIC controller by Microchip, which is well supported by a development kit for CAN bus application.

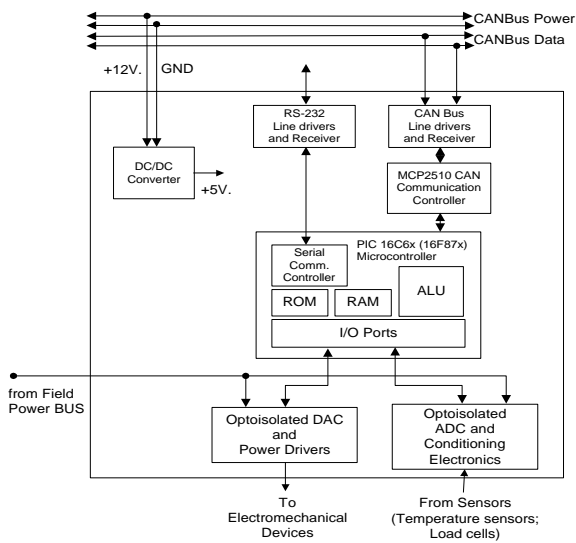

Figure 3: General architecture of the microcontrolled modules

The PIC component family chosen for the application development is the 16F87x, a flash memory microcontroller providing program memory range from 2k to 8k bytes. A component with flash memory is employed during the development activities because it allows for fast programming (no erasing process is needed) and the use of the in-circuit debugger, [3]. Both aspects speed up the development time. The implementation of the final version of the firmware will use the 16C6x OTP (One Time Programmable) microcontroller family. This family is equivalent to the flash component. The OTP components are cost effective and less subject to memory erasing problems.

All the firmware is written in ANSI C language.

A general architecture/diagram of a microcontrolled module for CAN bus is shown in Figure 3.

The microcontroller integrates an ALU, ROM for firmware storage and RAM for program execution. The CAN bus communication controller for protocol management is the MCP2510 by Microchip. This controller implements full CAN 2.0 A and B, at a transmission rate of 1 Mbit/s. The following components are also present in the module: CAN bus and RS232 line drivers, opto-insulated driver for motors activation, opto-insulated serial ADC and conditioning electronics to interface external analog sensors such as the load cells.

## 3  REFERENCES

[1] Mancini, D., Sedmak, G., Brescia, M, Cortecchia, F., Fierro, D., Fiume, V., Marra, G., Perrotta, F., Rovedi, F., Schipani, P.,: 2000, "VST project: technical overview", in "Telescope Structures, Enclosures, Controls, Assembly / Integration / Validation, and Commissioning", eds Sebring, T., Andersen, T., SPIE, 4004, 79

[2] Fantinel, D., Bortoletto, F., Pernechele, C., Gardiol, D.: 2000, "TNG: the active optic system", http://www.tng.iac.es

[3] Mancini, D., Sedmak, G., Brescia, M, Cortecchia, F., Fierro, D., Fiume Garelli, V., Marra, G., Perrotta, F., Schipani, P.,: 2000, "VST Final Design Review", http://vst.na.astro.it